# Triplication: an important component of the modern scientific method.

**Authors:** Jeremy S.C. Clark[1*], Karina Szczypiór-Piasecka[2], Kamila Rydzewska[1], Konrad Podsiadło[1].

**Affiliations:**

[1] Department of Clinical & Molecular Biochemistry, Pomeranian Medical University, Szczecin, Poland.

[2] Department of Rehabilitation of the Musculoskeletal System, Pomeranian Medical University, Szczecin, Poland.

***Corresponding Author:**

Jeremy Clark. Email: Jeremy.Clark@pum.edu.pl; Tel. 0048914661490;

**Orcid:** JSCC: 0000-0003-3179-5638; KS: 0000-0002-9562-9201; KR: 0000-0001-7416-0929; KP: 0000-0002-4061-2103.

**Short title:** Triplication in scientific methodology.

# Abstract.

A scientific-study protocol (defined) is designed to deliver results from which inductive inference is allowed. In the nineteenth century, triplication was introduced into the plant sciences and Fisher's $p<0.05$ rule (1925) incorporated into triple-result protocols designed to counter random/systematic errors which contribute to real-world variability. The aims of the present study were to: (1) classify replication protocols; (2) assess their prevalence in plant-science studies (published during one twenty-first-century year; for defined variable construct); (3) explore triplication rationale. Methods: a plant-sciences protocol-prevalence report was produced; experimental/associational-study proportions analyzed; and real-world-data proxies used to show confidence-interval-width patterns with increasing replicate number. Results: 25% plant-science studies analyzed showed triplication, including 11% triple-result protocols (including greater replicate numbers: 48%;17%, respectively). Theoretical considerations indicated that even if systematic errors predominate, (previously-known) square-root rules sometimes apply, contributing to triplication importance (exemplified by real-world-data proxies). Conclusions: The defined protocols, with minor modifications, should provide the means for assessment of most sciences. Triplication was extensively applied in studies analysed and there are strong methodological reasons why triplication, rather than duplication/quadruplication, is the appropriate standard: triple-result protocols: (a) effectively reduce false positives to acceptable levels; (b) give qualitatively-different information (shape) from duplication; (c) have a large efficiency advantage (concerning confidence-interval widths) over quadruplication. The application of batch replication is not, primarily, a statistical problem and cannot effectively be replaced by simulation.

# Introduction.

The fundamental methodological approaches by which hypotheses are tested and validated, including chosen replication model (here referred to as "protocol"), vary considerably across and within scientific disciplines. This often lacks clear justification, despite the fact it impacts on whether inductive inference can reasonably be made for the purpose of further hypothesis generation.

Given known, previous, use of triplication in the plant sciences, this study aimed to estimate the prevalence of replication protocols, including triplication, across the discipline during a single year in the twenty-first century, focusing on a specific combination of variables (referred to as a "variable construct"). The present study also sought to explore the theoretical foundations for triplication protocols.

Biological-batch replication (hereafter referred to as "replication" unless otherwise stated) probably always involves intentional or unintentional resetting of systematic errors arising from both known, unknown, potential or unforeseen, confounders. Replication, here, refers to several batches of a study's lowest-level "subjects" and does not describe subject numbers. Replication here should not be understood as mere continuation of data collection under identical conditions. Although the latter might result in lowering the (p-value) error rate due to random error alone, it ignores the fact that systematic errors might affect the whole dataset, i.e., that real-world variability does not only involve main-effect-parameter random error. In a resetting (see file_S1) replication, as many possible sources of systematic errors are reset between replicates as advisable; these decisions being difficult and subject-dependent.

Systematic errors have been precisely defined: Norena et al.[1] proposed a mathematical framework with some systematic-error components associated with probability-density functions and totals estimated via quadratic addition. They stated that nuisance parameters can be identified by uncovering science "that is invariant under arbitrary rescalings of the nuisance quantities" (in truth: identification of important confounders might never be achieved). Simpler definitions are ignored, e.g., an offset which is “constant” or “always in one direction”, preferring a working definition in which they only usually, on average, offset a result. It follows (and also from simpler definitions) that many systematic errors have their own random component, not precluding some having constant values in particular replicates. A positive result can therefore (by definition) only arise from (1) a true group effect; (2) main-effect-parameter random error; (3) systematic errors with random-error components; or (4) any combination of (1):(3).

In the present article, a "confounder" (=nuisance parameter) can give a systematic error (=bias) which sometimes results in a false positive, independently from main-effect-parameter random error which can also give a false positive. The most important role of any protocol, and the scientific method in general, is false-positive elimination; including efficient removal of false positives due to unknown confounders, the ubiquitous presence of which appears to be a fundamental aspect of reality[2].

Here, a "systematic error" can affect a datapoint or mean; or can arise from combinations of (known, unknown) systematic errors. Intra-replicate systematic effects can contribute to inter-replicate systematic errors, often uncovered by the replication resetting process (but not usually distinguished here). There are few general articles on systematic errors (but see examples in: taxonomy[3]; metabolomics[4]; evolution[5]; biomechanics[6]; genome-wide-association studies[7]); although in the plant sciences this might be because statistical treatment is already well-established[8, 9]. The reader is advised to read Supplemental_file_S1_Definitions before continuing.

Supplemental_file_S2 gives 1980s/1990s plant-science examples (n=100) with triplication (mostly uncharacterized). Nine examples employed "replication-error protocols" in which the variation among replicate means ("replication error") was reported without p-value statistics. Notably, these focused on direct assessment of (primarily) inter-replicate systematic errors, in contrast to studies which evaluate (primarily) random error using non-replicated p-values. Cross-talk between these two methodologies occurs (and, obviously, they could be used simultaneously): at large sample sizes non-replicated p-values will respond to some systematic errors, and at low sample sizes replication errors will respond more to main-effect-parameter random error. Generally, in a non-replicated study, p-values tend to change with increasing sample size primarily due to random error. In contrast, replication errors tend to vary with the number of replicates, mainly reflecting systematic error. When each replicate has large sample size (n), replication errors can reflect stable systematic differences between replicates and introducing a new replicate may cause a substantial p-value shift, highlighting possible systematic effects.

The implicit assumption in the nine examples above appears to have been that systematic error is always present and, as positive results were inferred after replication-error evaluation, likely had no meaningful effect. These researchers (or the journals) may have considered systematic error to be of greater relevance than random error (as indicated by the lack of p-values; note some plant journals do not allow p-values)—a perspective perhaps more broadly applicable.

Examples from the field of physics are included here because they provide precise, quantitative estimates of (usually known) systematic errors reported alongside results. For instance, the estimated age of the universe has been expressed as [10]:

$$\mathbf{t_U = 13.5^{+0.16}_{-0.14}\ (stat.) \pm 0.23 (or\ 0.33)\ (sys.)\ \ billion\ years,} \qquad \mathbf{(1)}$$

where: stat. = random error, sys. = systematic error.

In 25 reviewed physics studies here, systematic error was usually of the same order of magnitude or greater than random error.

In the medical sciences it is easy to show that false positives due to systematic errors occur ~10×:15× more often than due to random error: from [11]: false-positive prevalence was 0.5:0.9; from [12]: systematic-error false-positive prevalence = "false-positive prevalence minus 5%"; as a ratio this gives $\frac{(0.5:0.9)-0.05}{0.05}$ = 9×:17× random-error false-positive prevalence.

Triplication has a long history in the plant sciences, with examples from the 1890s: in Escombe[13], p. 587; triplication was recorded from Peter[14]; from Bedford[15]: "...were sown in triplicate..."; Keeble[16]: "...three sets of leaves..."; and the U.S. Department of Agriculture: "Three distinct samples were taken in this way from each plat…"[17]; "...clover seeded in rye on triplicate plats..."[18]. Amid ~14 years that Fisher worked in the plant sciences at Rothamsted Experimental Station, UK, and after he published the 1925 p<0.05 rule[19], he might well have expected that optimal science would utilize a triple-result protocol, i.e., p<0.05 three times. In his 1926, 1929 and 1935 papers, Fisher affirmed that it was important to establish replicated results alongside the p<0.05 rule[20–22]. Around that time, several examples of triple-result protocols appeared: e.g., Preston & Bragg (1934[23]; Table II) reported three independent measurements with p<0.05; Lamm (1936[24]; Table 6) presented three plant population samples with p<0.05; Thomas and Hill (1937[25]; Table 4) provided triplicated results. In later years, one of the present authors (JSCC) reports having applied part-study triplication extensively between 1985 and 2000 across three plant science institutes in two countries, encompassing nine different projects.

Discussions of p-values concluding that p<0.05 alone is unacceptable as a means to fully test a hypothesis are given by Benjamin et al.[26], Randall and Welser[27] and Fisher[21]. Benjamin et al.[26] suggested lowering the significance level to 0.005 for a non-repeated investigation and Randall and Welser[27], in a replication network publication (https://replicationnetwork.com), stated that "p<0.01 should be the loosest recognised standard of statistical significance, not the most rigorous".

A triple result is, however, far stronger than either of these if positive coherence across replicates is a requirement (three replicates individually significant, p<0.05). Firstly, this provides a stronger counter to random error:

$$\textbf{Triplejoint_p} = \mathbf{0.05^3} = \mathbf{1.25 \times 10^{-4}} = \textbf{Fm_altered}(\boldsymbol{p_{1:3}} = \mathbf{0.05}), \qquad (2)$$

assuming joint probability under independence with each replicate unbiased [28, 29]; in general, a product of any p-values (joint_p) equals Fm_altered(p-values), where Fm_altered = (Fm with only alteration of constant 2 degrees of freedom (df)), see below (Supplemental_file_S11).

In contrast, Fisher's method for combining p-values with aggregate null [30]:

$$\textbf{Fm} = \chi^2\left(-2 \cdot \sum\left(\log_e(\text{p_values})\right), \textbf{df} = 2n = 6 \textbf{ for } n = 3 \textbf{ replicates}\right) \qquad (3)$$

uses a p<0.05 cut-off, allowing non-coherent aggregate extremeness to become significant. This is designed for combining several independent experiments with none individually conclusive and has consequences concerning inter-replicate systematic errors. It allows three non-significant results ($p_{1:3} = 0.12$: Fm($p_{1:3}$) $= 0.048$) or only one significant result ($p_1 = 0.007$, $p_{2:3} = 0.5$;

$\mathrm{Fm}(p_{1:3}) = 0.048$) to pass overall significance, a considerably weaker procedure not acceptable for most studies unless vulnerability to corruption by a systematic error in just one replicate is ignored, whereas a triple-result-coherence requirement resists this vulnerability.

($\mathrm{Fm}(p_{1:3} = 0.05) = 0.0063$: not used in decision-making here.)

In either case, the presence of inter-replicate systematic errors (introducing "calibration" bias to individual replicates) would, theoretically, demand a lower (joint) cut-off to restore the intended false-positive rate; and shared bias (i.e., dependency; not countered by replication) even lower [31, 32]. In the interests of standardization a combined triple-result significance level can be taken, nominally, to be $p < 1.25 \times 10^{-4}$ (with coherence requirement (p<0.05 three times), no calibration bias and independence assumptions above; empirical estimates possible; file_S11).

Secondly, inter-replicate systematic errors, more likely to give false positives than random errors (at least in the domains reviewed here), are more effectively addressed through resetting replication protocols, including a triple result. A low p-value obtained from a non-replicated study provides no information about such errors.

Discussions on p-values, while important, might distract from the (likely) more important systematic errors. A p-value often only applies to one replicate (or rather an imaginary collection of replicates of identical characteristics) or to data collected further without resetting. When a second replicate is assembled independently of the first, real-world variability (with effectively infinite degrees of freedom) consistently introduces differences not captured by theoretical assumptions. Even after following identical written protocols, systematic effects are likely reset: hence real-world replication may (and often does: see later) differ meaningfully from hypothetical, idealized or simulated, repetition.

The common pattern of a positive followed by a negative result often (probably most likely, given the calculation above) indicates the initial finding was a false positive caused by (unknown) systematic errors. While p-values may appear consistent across replicates in some studies, this is not a universal rule. In fact, in several scientific disciplines replication is considered essential for this very reason, with the motto: "no replication, no result".

To establish optimal procedure (and the state of a science) it is important to assess protocol prevalence as well as assessing rationale. The present article has therefore assessed the prevalence of defined protocols in the plant sciences which use p-values or replication errors to establish a result.

Protocol characterization involved considerable technical complications (see Methods). One year (2017) was chosen; 100 articles assessed; because: (1) protocol reports take a great deal of preparation time; (2) a power study showed sufficient power; (3) future studies of other years might indicate trends; (4) the date chosen was close to that of Freedman et al.[11], used in the production of the weighted directed graphs; and (5) more recent articles are more likely paywalled.

The main aims of the present study were to assess replication-protocol prevalence across the plant sciences from a twenty-first century year, and to describe the triplication theoretical basis with support from real-world data. The hypotheses were (1) protocol prevalences have a distribution which can indicate the importance assigned to plant-science replication; (2) there is a distinction between associational/experimental studies in terms of different protocol use; (3) real-world data follows theoretical patterns created concerning pooled confidence intervals (CIs). The main protocol report question was: from an article selected from the plant sciences, which of the protocols (defined in Methods) were used for a study result of the relevant variable construct (see below).

# Materials and methods.

## Study design.

A protocol report was prepared based on plant-science articles published in 2017, covering almost (see exceptions below) the entire field. Key decisions included: (1) selecting journals; (2) inclusion/exclusion of articles; (3) identifying a specific study to analyze within each article; (4) classifying the replication protocol; (5) classifying the study as experimental/association-based; and (6) estimating potential risks associated with decisions (3):(5).

Further studies involved using proxy data to show the effects on confidence-interval widths (CIwidths) with increasing numbers of replicates.

## Relevant variable construct.

The protocol report in the present article was restricted to studies discerning continuous or ordinal main-effect-parameter differences between categorically-distinct groups. This defined the "relevant variable construct" and major conclusions from the present study only apply if one replicate can be presented as in Fig 1.

Usually, but not necessarily (replication-error studies providing exceptions), frequentist statistics with p-values were applied, and significant interactions/effects due to replicate were found to be absent from presented results.

It was imperative to restrict the variable construct to avoid complications regarding internal corroboration. Notably, groups defined ordinally or correlations between continuous parameters were not allowed. (Replication also applies to other variable constructs but conclusions might differ regarding triplication.)

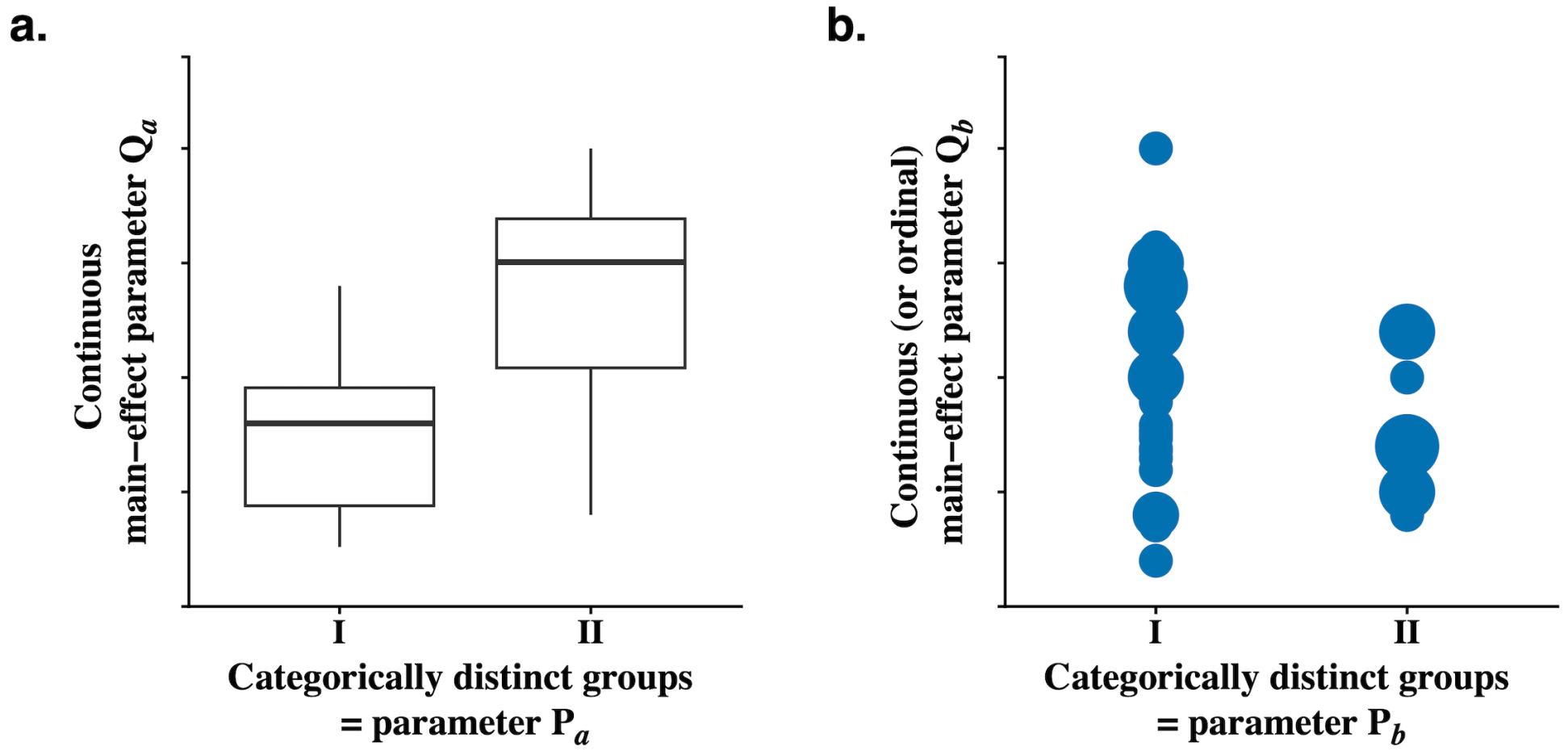

**Figure 1. Relevant variable construct.** Graphical examples of an individual replicate with the "relevant variable construct" applicable to the present article, with two groups I and II. The variable construct had to have at least two categorically-distinct groups (giving parameter P) with enough values in each group to test differences between the groups; an unlimited number of further categorically-distinct groups could also have been deployed. The main-effect-parameter (Q) real-number (continuous or discrete, e.g., integer) or ordinal values could have any distribution. No Q value could be a statistically- or mathematically-derived modeling parameter from an organism series (e.g., from regression, bulk segregant or aster analyses). (Proportions/amounts from pooled binary outcomes from biological replicates, or direct physical/chemical measurements from pooled biological material, were allowed and were not counted here as derived parameters.) For conclusions from the present article to apply, a positive result indicating at least one difference among the groups had to be deduced statistically, or shown using replication errors or simple variance errors, and it had to be possible to present an individual replicate as in a.: box-midlines represent medians, box edges: interquartile ranges; whiskers: ranges; or b.: dot areas represent numbers of binned values.

## Protocol report.

Full journal and article search strategies are given in sections below (in file_S3_PRISMA and in the heuristic guide in Supplemental_file_S13). In brief, all journals were considered for the protocol report from the Web of Science (WoS) from the section 'Plant Sciences'. Plant science does not only involve the study of plants (see file_S1_Definitions), and studies were included of any organism found in the journals (e.g., fungi, algae, bacteria) across the entire area of plant science apart from, perhaps, some areas of ecology or mycology or forestry (when indicated by the WoS categories "Mycology" or "Forestry"; see below); non-biological results excluded. Additionally, the actual "subjects" in a plant science study vary considerably (e.g., a microscope field, a pooled juice sample or a tree etc.; a longer list is given in S1_Definitions:subject). We have come across no indication whatsoever that different replication protocols are consistently applied to different plant science sub-categories or topics apart from that of quantitative RNA work (excluded; possibly five replicate studies might be demanded). There is no other reason known to us for any sub-division of the field of plant sciences (even the separation from WoS categories "Mycology" or "Forestry" is likely unnecessary in this regard) and protocols demanded by journals or Institutes, as far as we know, vary little according to sub-discipline. Ecology journals were not included (several have large numbers of animal studies); but from included journals no distinction was made between ecological and other studies.

Initially, an impact-factor (IF) grid was constructed from the Web of Science to ensure balanced representation across IF ranges (although subsequently the grid became superfluous). From selected journals, articles were selected using a randomization technique and were assessed for protocol-count eligibility. One study from each article was classified according to which replication protocol (defined in Methods:Definitions) was determined to have been used from replication groups: 1) non-replicated studies; 2) global protocols (usually with "method replication"); or 3) replicated-result protocols. All studies had to have p-values, replication errors (which are errors across batches in a replicated protocol; file_S1_Definitions), or non-overlapping simple variance errors at the lowest "subject" level (defined as either standard errors, variances or sds describing single-sample variance in a non-replicated protocol), showing differences between categorically-distinct groups with the relevant variable construct (Fig 1). A triple-result protocol meant that an investigation had been performed with three sets of plants, fungi or other organisms, and that either all sets (or an individual representative set) of results were published.

The final number of studies chosen to be selected was 100, as protocol-report analyses took considerable time and a simulation-based power study indicated this was of sufficient size (see below). The protocol report question is given in the Introduction and involved a search for the protocols defined below.

## Electronic sources.

Electronic sources were: Journal Citation Reports (Clarivate; https://wokinfo.com; Web of Science; WoS), 2017 list; date first searched: October 2018; data last searched for corrections: March 2026, initial selected categories:  "Forestry", "Mycology", and "Plant Sciences"; Google

(https://scholar.google.com). Only journals with WoS listing as "Plant Sciences" were subsequently included.

## Journal search strategy.

A journal grid was created from the 2017 Web of Science JCR list (file_S4) SCIE, SSCI Categories. Included journals were published in English, and initial journal search start points were used in a grid across impact factor ranges. However, subsequently this resulted in all plant science journals assessed above IF1 (the grid therefore becoming superfluous), and a further twelve articles were taken by assessing decreasing impact factors from IF1 to reach the target of 100 included articles. Exclusion criteria for journals/issues were: ecology, taxonomy, methods, economics, genetic diversity, medicine, pharmaceutics, or biotechnology, or without Web of Science (WoS) category "Plant sciences", or without relevant variable construct, not primary research, or not available. In practice if a journal name contained, e.g., "ecology" in the title, this was automatically excluded. With doubt, journal exclusion occurred when articles were "mostly" excluded,  "mostly" defined as when, on first encounter with the journal, the first 10 articles followed an exclusion criterion. The full list of excluded journals, together with reasons, are given in file_S4 (articles from excluded journals are given in Supplemental_file_S12).

Journals were included if they contained articles which conformed to the article inclusion criteria (given below), with primary original research, full articles available in English, without rare subjects. Note there were missing outcomes because journals not freely available were not covered by the assessment.

## Article search strategy.

Articles were included if they (1) had a study with the relevant variable construct; (2) had at least one positive result defined by a statistically-significant p-value comparison, by replication errors or by non-overlapping simple variance errors (defined in file_S1); and (3) did not fulfil article exclusion criteria (see below). All included and excluded articles (with exclusion descriptions) are given in Supplemental_files_S4,S5.

Articles were always from different journals (S4) and were included or excluded according to the presence or absence of the relevant variable construct, but exclusion descriptions were given as: -omics studies (e.g., genome-wide association studies, RNA expression arrays), studies on genetic diversity, taxonomy, population structure, selection, or evolution. These descriptions described the main reason, very often, why a relevant variable construct was absent. In practice, 10 articles from each journal were surveyed for variable constructs (Annali di botanica only had eight 2017 articles), so main exclusion criterion was lack of relevant variable construct. Also excluded were: pilot studies, reviews, simulations or modeling papers, technical-method reports, or with rare subjects, as well as cell studies from multicellular organisms.

A complication arose for genetically modified organisms (gmos): standard production uses three independent experiments producing only three individual organisms (and therefore excluded as "rare"). However, later, each line might have many plants/individual organisms, and it is a difficult question as to how many individuals, how many lines, and how many repetitions of an experiment should be included in order to sufficiently account for biological variation. Here, gmo studies were only included if the gmos had been produced in a previous study or if they had been bulked within the article. (All unicellular gmos were assumed to have been bulked and were included.)

Exclusion did not mean inferior comparisons (e.g., categorically-distinct groups repeated across a concentration gradient would usually give a superior variable construct). Conversely, a non-replicated result necessarily had to have a positive result demonstrated either statistically or by non-overlapping simple variance errors at the lowest "subject" level, which excluded many results which might otherwise be considered as non-replicated (but were here considered as descriptive).

If no suitable articles were found within 10 articles, then the main reason was noted in the grid and the next journal was assessed. (Above impact factor 3.7, many articles were multifaceted with many studies: only one study was selected.)

Article selection (with random start points: random numbers created using R function *stats::runif()*) was carried out by three authors (JSCC, KP and KR by joint decision), and checked later by JSCC. Disagreements were resolved by JSCC. No data were extracted from overlapping or duplicate reports. The risk of bias in article selection was thought to be minimal as the inclusion criteria included articles with a wide range of possible sub-disciplines. The risk of uncertainty in decision-making was given for each article (file_S5) and included: 1) Risk: Main result = risk that comparison was not for main result. 2) Risk: Repetition = how unclear was the number of times a study was replicated; 3) Risk: Repetition of p-value = how unclear was the number of times p-values were replicated (e.g., to decide between a triple-result or global protocol); and 4) Risk: associational/experimental = reflecting doubts concerning identification of study as associational or experimental.

## Technical complications.

Sometimes an entire study was repeated three times but only one pooled result given; or three results; or one individual representative result. Sometimes both pooled and individual results were given. Sometimes it was unclear whether the result was obtained once, twice, thrice or what, exactly, was triplicated: subject, part-study, or entire study. Sometimes global statistics were given, with or without individual results.

It was decided that, with clear presentation showing (1) triplication of entire study; and (2) individual results (or representative result) presented (regardless of pooled results); then "triple-result" or "result triplication" would be declared. With lack of clarity regarding triplication, a catch-all "global protocol", "method triplication" would be declared, reflecting this sometimes involved a global statistic following result pooling (and similarly for duplication).

Regarding replication levels: the lowest "subject" levels (defined here as non-batch replication or replication-level zero; see file_S1:"subjects") were excluded from (batch) replication definitions. At the highest level, if an experiment had only one instance, (batch) replication might, however, have occurred in parallel within this (Supplemental_file_S5).

## Replication and repetition definitions (see also Supplemental_file_S1_Definitions).

Concerning the results of analyzed articles, use of the words "replication", "duplication" or "triplication" apply to biological repetition unless otherwise stated. Replication could occur either in time (consecutively), space (in parallel) or with groups of organisms, in that the study or part-study for the comparison being evaluated occurred more than once, including a new set of plants (or other organisms).

Whether a study or part-study is replicated is relative to the hypothesis. For example, if a hypothesis involved investigating whether species differ, then analyzing three species separately would not be counted as "replication". However, if the hypothesis involved investigating a phenomenon for which marker species can similarly be used, then analysis of three species might well result in a "replicated" classification. Similarly, years or sites might give ordinal arrays or not, depending on the question being considered.

Exact replication (and pseudo-replication categories) are here not distinguished, i.e., systematic differences between replicates might or might not have been introduced with known confounders (this would be relative to the written methods). It can be assumed that (reset) replicates always have unknown (including potential or unforeseen) differences and it is quite possible that unknown-confounder effects often overwhelm known-confounder effects.

"Result replication", where results are given several times or an individual representative result is presented, is here distinguished from "method replication" with a "global" classification, in which there is clearly replication in the methods but results are pooled, often using a global statistic.

Of importance, if a study had several replication levels (e.g., non-batch replication: refering to multiple subjects at the lowest level plus one or more batch replication levels), then the article was classified according to the top level of replication. For example, if an entire experiment had been duplicated, and part-studies had been triplicated, then the article was classified as having duplication (see full descriptions of levels in file_S5). "Global protocol" was here used as a catch-all classification, designed to include all studies in which some form of (batch) replication had occurred, but without presented individual results. An extreme in this classification could include where an entire investigation was performed three times consecutively, but then only pooled results presented.

## Protocol definitions.

(A term with "+" is used to refer to that number or more of replicates).

### 1a. Non-replicated protocol (NoR).

Declared if no evidence of (batch) replication was discovered in a study with a positive result (shown by a p-value or other method, e.g., p-value limit). All studies assigned this protocol had no batch replication, but usually a lowest, non-batch, level with many subjects (referred to as a non-batch replication level zero or the lowest "subject" level). A dilemma arose concerning randomized block designs which often had no (batch) replication by this definition, with one lowest "subject" level, e.g., with only one plot per treatment per block. This was included here as NoR (assuming no other levels) and did not count as batch replication by this definition. The fact that blocks were used was regarded as a design structure rather than giving a replication level, even though the physical placing of the "subjects" across an array was likely to counter some forms of systematic spatial error (e.g., light, temperature, or watering gradients within a greenhouse), resulting in the dilemma. This was resolved conservatively in favor of NoR in order to align with other designs: for example a simple "subject" list might also counter some systematic errors depending on how the subjects were selected, but is not thought to be as effective as batch replication. Randomized block designs with two or more "subjects" per block were counted as batch replication and assigned another protocol (see "Global" protocols).

## 1b. Non-replicated protocol (N3).

As in 1a but with exactly three lowest level "subjects". Other numbers of lowest level subjects are also indicated (with NoR classification).

## 2a. Replication-error protocol (duplicated or triplicated+).

Declared if (batch) replication errors (standard errors, variances or sds among replicates) were found with no p-values given (or other measure, e.g., effect size), i.e., that random error did not have a direct assessment. (These involve pooling and therefore form a subset of global protocols; ordinal embedding was allowed - as in 3b.)

## 2b. Global (duplicated, G2; or triplicated+, G3+) protocol.

Declared if the study clearly had (batch) duplication or triplication+ in the methods, but no individual replicate results given, i.e., only global result p-values (or replication errors) presented. This was declared as default if doubts arose concerning whether individual result replicates had been presented. A global protocol can counter systematic errors in a study or part-study by indicating whether replicate has a substantial effect (or, with the biological definition (see below), systematic effects could be averaged by pooling data or samples from biologically distinct individuals). This protocol was sometimes difficult to distinguish from a non-replicated protocol.

Global protocols (G2/G3/G3+) were defined either structurally or biologically (made explicit in the Protocol Report):

## 2b.1. Global protocol: structural definition.

The Global protocol structural definition required physical batch replication above the lowest subject level. Randomised complete block designs with only one unit per treatment per block in a single experimental run would not count as batch replication by this definition (the block was regarded as a design structure rather than a replication level). However, randomised complete block designs with two or more units per treatment per block in a single experimental run did count as batch replication by this definition.

## 2b.2. Global protocol: biological definition.

With the Global protocol biological definition, within-unit sub-samples that contained biologically distinct individuals constituted sub-sample pooling, and the next level up (e.g., a number of units) provided batch replication by this definition. Note this could give a grey overlap with non-replicated protocols depending on how an individual is defined - and global protocol is a catch-all category which includes many ways in which pooling could occur.

A complication arose when within-unit sub-samples produced only binary outcomes that aggregated to a unit-level proportion (e.g., germination %, survival %): the proportion was counted as the lowest (replication zero) level and this was not allocated two (batch or non-batch) replication levels.

It was decided that if several measurements were made from one sample and these were pooled to give one pooled measurement for one sample, then this would give the lowest "subject" (replication zero) level, whether or not the pooled measurement was reported with simple variance errors. With another level of pooling performed on top of this, leading to a pooled pooled measurement, this would be counted as batch replication. In other words, the global protocol reflects the physical reality of the set-up, as well as statistical considerations.

### 3. Triple (or other multiple, T3+)-result protocol.

Declared if positive results were given for all individual replicates or for a representative replicate. Global results might have been presented, but individual results were also presented from one or more individual replicates from a set. The standard significance level for a triple-result protocol was usually $p<0.05$ three times and, therefore, random error was countered by an overall significance level of $p\sim<10^{-4}$ or less (see Introduction).

## Assessment of study protocol.

The assessment of study protocol followed the heuristic guide in file_S13 and, roughly, the columns given in Sections C1:C3 of file_S5. Firstly, for C1, a relevant aim was chosen ("Chosen relevant aim/result with categorically distinct groups"); relevant groups ("Categorically distinct groups"); a dependent parameter ("Chosen relevant quantitative parameter"); a positive result (e.g., "One relevant p value, limit or result"); the factor replicated ("Factor replicated"). For C2: the "(batch or non-batch) replication levels" were determined. For C3: evidence was found for method and/or result replication ("Method or Result Replication"); and for D: the risk. Finally, the experimental or association category was determined ("Evidence for manipulation or random placement of groups during study") and risk.

## Assessment of experimental or association study type.

The definition of experiment was strict, i.e., there had to be placement or severe artificial manipulation of the environment of subjects. Quasi- or pseudo-experiments involving random selection (but not placement) of subjects were not allowed in the category "experiment". The method involved finding evidence for random placement or severe manipulation of the relevant groups (file_S5:Section_E: "Evidence for manipulation or random placement of groups during study"). If none were found the study was classified as an Association study.

## Minor studies.

Supplemental_file_S2 gives 100 examples of triplicated plant science studies from the 1980s/1990s. These are not part of the protocol report and were found by searching plant science journals for the text "three". Journals were chosen from Web of Science (https://jcr.incites. thomsonreuters.com), impact factor list from 2018. Years were chosen nearest to 1980, 1990 and 1999 from each journal. Articles were included if they were full length original articles with "three" indicating number of replications of the entire investigation. Pilot studies, reviews, and modeling papers were excluded, as were studies in which replication was unclear (e.g., ambiguously stating: "mean of three replicates"). (This unclear exclusion criterion only applied to minor studies; in the Protocol Report this was often handled as increased risk with reason noted.)

Supplemental_file_S6 shows 25 examples from physics where systematic errors were quantitatively measured. These articles were the first found using a Google Scholar search for "physics" and "syst" and "systematic error" in October 2025.

## Assignment of relative CIs from real pooled data.

In order to show patterns in how CIs change with sequential addition of replicate results, the data used (with important limitations discussed) were from all studies initially classified as triple+-result studies in the protocol report (plus previously unpublished data to fill an otherwise missing space in Fig 4). Data was extracted from articles: A. Tonk et al. [33]; B. Kichigina et al. [34]; C. Carlucci et al. [35]; D. Lawrence et al. [36]; E. Zhang et al. [37]; F. Driver et al. [38]; G. Zhou et al. [39], and H. unpublished data from a study preliminary to Clark et al. [40] (which contains isolate details). Details of extraction methods for all data are given in Supplemental_files_S7,S8. The data were used to simulate the effect of adding replicates sequentially on resulting CIs, with or without systematic-error estimation, for pooled data from a single group/parameter. Means and sds/standard errors from replicates were taken in the order reported; the first replicate representing k=1 and subsequent replicates pooled consecutively to generate values for k=2 onward.

CI-component estimates from random error used pooled degrees of freedom ($df = \sum(n_i - 1)$) ; Welch–Satterthwaite approximation was also applied as a sensitivity analysis (with limitations for combined uncertainty components discussed by Ballico [41]; see file_S7:Figs S7.B1-B2; TABLEs S7.T14 and S7.T17; file_S11:TABLEs S11.T5-T9). Systematic-error components were calculated as the standard error of the grand mean estimated from the variance among the means.

The CIwidth patterns (actual versus expected) with increasing replicate number (k) were compared in three ways (note these are all descriptive and the two patterns are not entirely independent): (1) using generalized additive (GAM) models with penalized splines (R mgcv::gam(s()) followed by stats::anova(test = "F"); (2) using observed-to-expected ratios $r_k$ (each width taken relative to k=2, so that $r_k$=1 at k=2 by construction) evaluated against both a ±25% multiplicative tolerance band (tau=1.25 giving tolerance 0.80:1.25) around the expected curve at all informative steps (k≥3; k=1 being the unpooled start and k=2 the anchor) and (3) evaluation against a ±25% additive tolerance band (actual minus expected as a percentage of the k=2 CI expected width). The tolerance bands were regarded as more informative for these data than GAM-based significance testing of observed-versus-expected differences, although the latter provided good descriptive curves (file_S11:Fig_S11).

## Assignment of relative CIs to theoretical pooled data.

A graph of theoretical (or "expected") changes in CIs was also created, assuming equal sample sizes and variances. The effects of adding systematic error (which followed a normal, t or uniform distribution) were also shown. All coding is given in Supplemental_file_S7 (which operates on Supplemental_File_S9).

## Lowest "subject" level numbers.

While not the main focus of the article, the numbers of lowest-level subjects were counted for all NoR/N3 protocols (file_S7:Section_6). Additionally, in file_S11:Section_2.2, means were treated as individual datapoints in order to illustrate the descriptive use of a normalized Greenwood statistic (Gw; [42]) which evaluates the variance of the gaps between ordered datapoints. Greenwood $= \sum p_{\mathrm{i}}^{2}$, where $p_{\mathrm{i}} = g_{\mathrm{i}}/\mathrm{range}$ and $g_{\mathrm{i}}$ is a difference (=gap) between two datapoints (for n values there are n-1 gaps), here normalized to give $\mathrm{Gw} = (\sum p_{\mathrm{i}}^{2} - \mathrm{floor})/(1 - \mathrm{floor})$, where $\mathrm{floor} = 1/(n-1)$ which is the minimum Greenwood value at a particular n. Gw's range is 0:1 whereas, e.g., a non-normalized Greenwood statistic for n=3 would range from 0.5:1, or for n=4 from 0.33:1.

## Reporting and statistical analyses.

The protocol report followed PRISMA guidelines (file_S3). The exact protocol-report question addressed is given in the Introduction. Note that this was a results-based assessment (not an assessment of stated intent) and, as far as we know, no review protocol exists for such a report.

All Supplemental files, including coding files, are found at https://github.com/Abiologist/Triplication.git or via http://arxiv.org/abs/2201.10960.

All statistical tests and all graphics (except the weighted directed graphs) were created using the R statistical platform (version 4.6.0; [43]; https://cran.r-project.org) on a MacBook Air (Tahoe 26.4) and all coding is given in Supplemental_files_S7,S11. (Inkscape version 1.4, www.inkscape.org, was used for the weighted directed graphs.) Claude Opus 4.8 (Anthropic, 2026), a generative artificial intelligence tool, was used for administrative purposes (especially with the Protocol Report).

An a priori omnibus power study (S7_Section_1B) was performed to estimate the number of articles needed by assuming a Poisson power-law null model (i.e., a smooth log-linear decay with increasing replicate number; the null assuming the population of studies is heterogeneous in cost-benefit tradeoffs) with one multiplicative bump placed at a single position as the alternative. Assumptions were: 90% studies with 1:4 replicates; r (baseline slope parameter) = 0.6 (sensitivity to r subsequently tested). Pearson chi-square departure from fit rejected at alpha = 0.05 (with $10^5$ simulations, using stats::rmultinom).

For collated data (S7_Sections_2,3) Poisson generalized linear models with a log link (i.e., power-law decays for count data) were fitted, with or without removal of n=3, for a quasi-Poisson F-test (robust to mild overdispersion) to exact_methods data (data shown in S7_Section_3A: this data included all replication numbers but with global and result replications not separated). S7:Section_3B_Fig_S7.A shows the log-log plot. Benjamini-Hochberg False Discovery Rate was applied to a quasi-Poisson scan of the first four points (S7:Section_4).

Data concerning non-batch replication is shown in S7:Section_6. For association-to-experiment ratios, Fisher's exact tests were applied (stats::fisher.test) and Cohen's W (rcompanion::cohenW(type = "bca")) calculated as standardized effect size. Two one-sided z-tests for equivalence were also performed. For one proportion (file_S7:Section_6), Wilson's CI was calculated with continuity correction (cardx::proportion_ci_wilson(correct = TRUE)).

Statistical significance was defined as p<0.05; all tests had two-tails; all simulations used $1 \times 10^5$ trials; all CIs were set at 95%; and power studies used alpha=0.05, power=90%.

# Results.

## Batch replication.

For 100 articles, a power study (chi-square departure from Poisson; single bump alternative) showed >90% power to detect 2.6× excess at n=3 (file_S7: other bump positions also assessed).

In 2017, plant-science replication was highly prevalent (Fig 2): from 100 articles, only 30% studies lacked replication (Fig 2:NoR; or 23% excluding N3); 25% studies used triplication (including 11% triple-result protocols); 48% had triplication+ (including 17% triple-result+ protocols). Using a quasi-Poisson F-test, n=3 study numbers were substantially and statistically-significantly increased over expected (2.65×, 95%CI:1.45×:4.58×; F-test p=0.012; partial D-squared=0.559; Fig 2; file_S7:Fig_S7.A). This category accounted for ~56% deviance unexplained by the Poisson fit. For this F-test, triplication was the a-priori hypothesis and the n=3 contrast specified in advance, not selected as the largest deviation. Comparisons among n=1:4 were also of interest, and a scan showed statistical significance survived Benjamini-Hochberg false-discovery-rate correction (file_S7).

This might, however, be easy to misinterpret. Although the largest replicated group (Fig 2: G3T=G3 or T) represents triplication, this does not necessarily mean results were triplicated (T); it means in some methodological part batch triplication occurred (with possible pooling). G3 is a catch-all protocol group including many ways in which triplication could be incorporated (see Methods). (Triplication decisions were usually relatively easy and quoted evidence is given for every article; file_S5).

Another complication was that, in 7% studies, the actual lowest-"subject" level (possibly substantial, e.g., a field) had 3 instances (N3; no batch replication). These were, nevertheless, categorized conservatively as non-replicated. (In the following, a plus sign after a number word/symbol indicates ≥: e.g., three+ means ≥3.)

Triplication+ occurred in 48% studies (Fig 2: G3T to GT6+) or, including N3, 55%. Additionally (Fig 2A; result replications), in 17% studies results were triplicated+. This means, despite immense publication pressures, these scientists spent two-thirds or more of their time, money and other resources (used for those results) on result triplication+ before publication.

There was a large decrease from G3T (25%) to GT4 (5%).

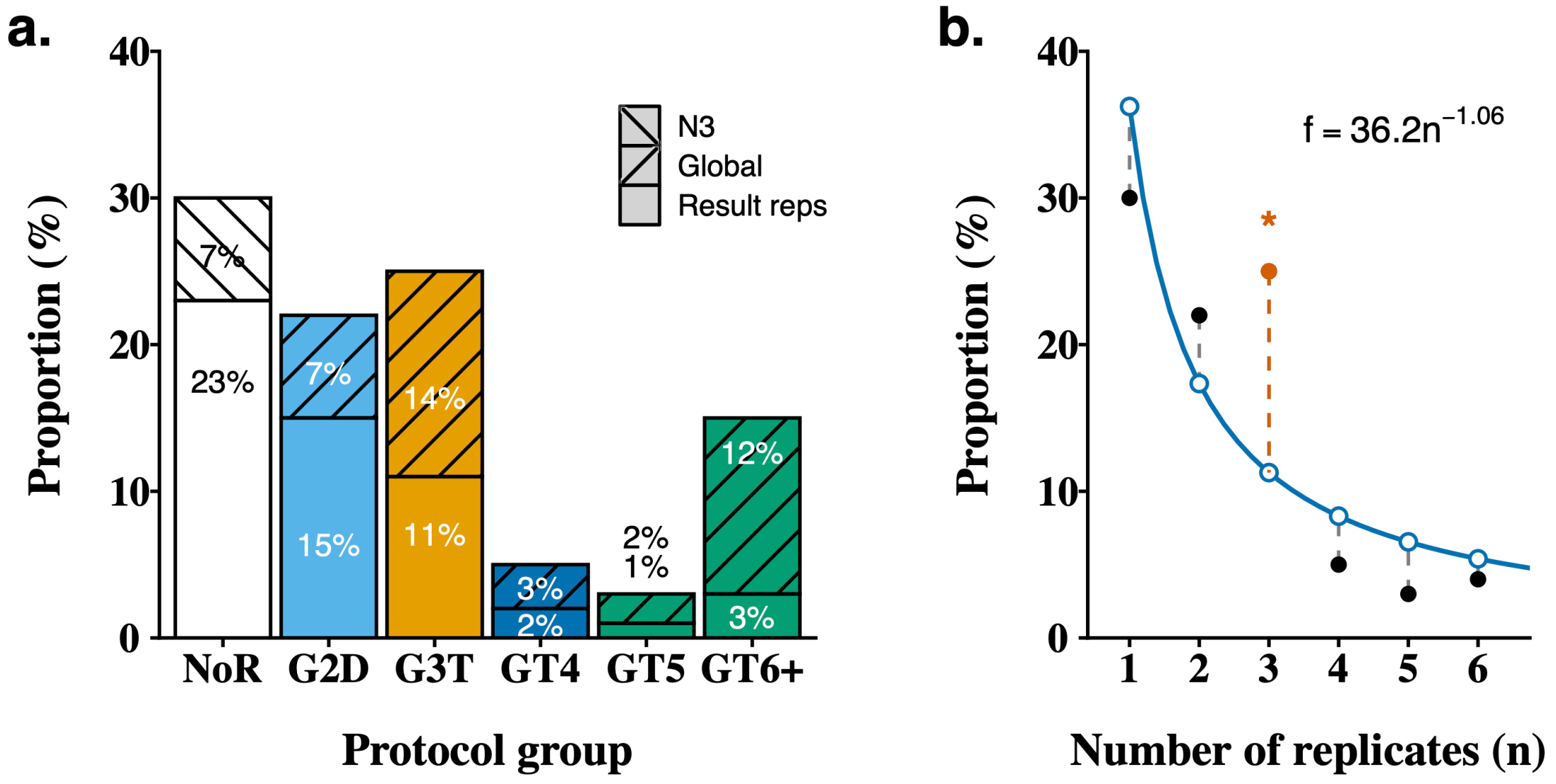

**Figure 2. Protocol proportions. 2A.** Proportions of studies (n=100) using the replication protocols defined in the text: NoR: no replication (N3 is a subset with exactly three lowest level "subjects"). Protocol bars with replication are stacked with Global and result replications (reps). G2D/G3T = global protocol with duplication/triplication in methods (G2/G3) or two/three result replicates (D/T), respectively; GT4/GT5: global protocol with four/five replicates in methods (G4/G5) or four/five result replicates (T4/T5), respectively; GT6+: global protocol with six or more replicates in methods (G6+) or six or more result replicates (T6+). **2B.** A Poisson power-law fit (f) to the same data (with GT6+ split into further categories; only 1:6 replicates shown). The value for three replicates is statistically-significantly (2.65×; quasi-Poisson F-test p=0.012) higher than a curve fitted to all data bar n=3 (note: the curve shown is fitted to all data).

## Experimental/associational-study type.

Experimental/associational-study type had little effect on applied protocol (Fig 3). For two groups: non-replicated/replicated, proportions (non-replicated: association = 6/16 = 0.38, experimental = 24/84 = 0.29) were not statistically-significantly different (Fisher's test: n=100, p=0.554, Cohen's W=0.0714, CI:0.00232:0.244). An equivalence test was not able to demonstrate difference/equivalence at this sample size (file_S7), but W=0.0714 is negligible: Cohen's [44] small-effect benchmark is 0.10 and the upper-confidence limit (0.244) was below the medium benchmark: 0.30.

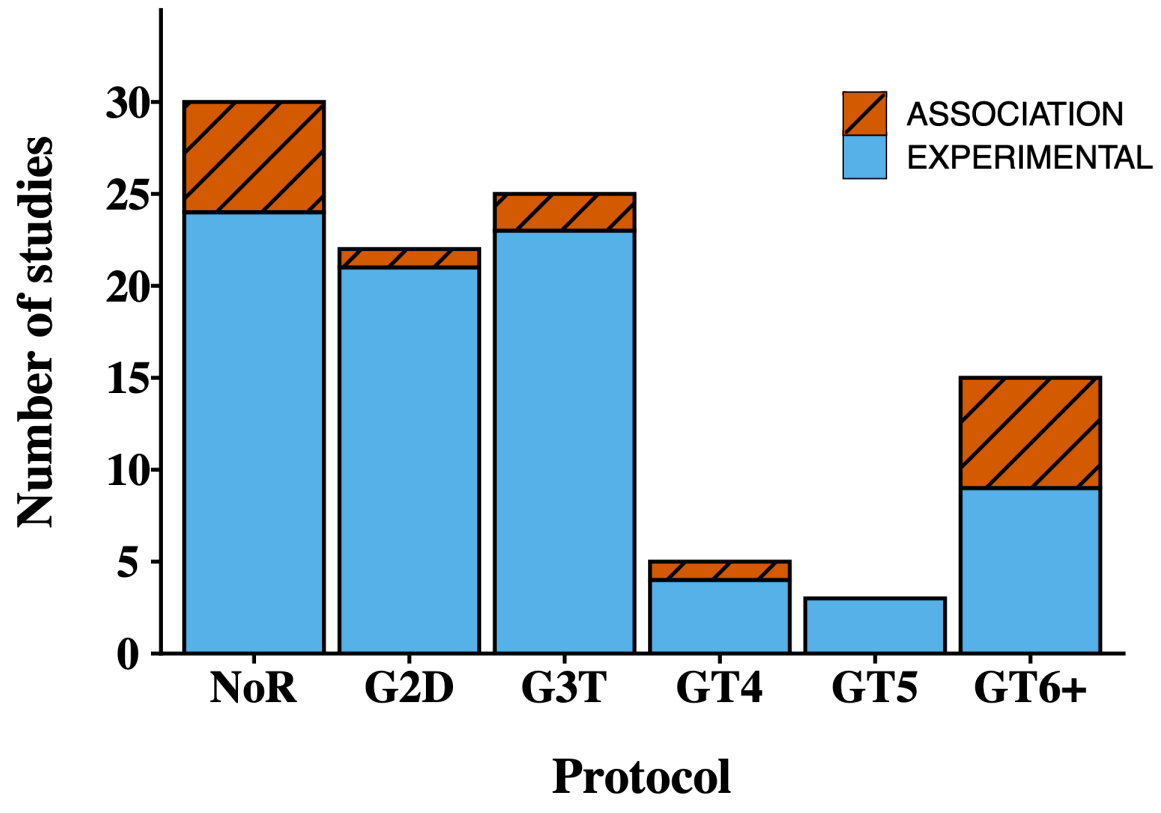

**Figure 3. Association or experimental study types.** Numbers of studies using the replication protocols defined in the text, classified as either an association or experimental study. It is clear that whether the study was associational or experimental made little difference to replication number. For protocol abbreviations see Fig 2.

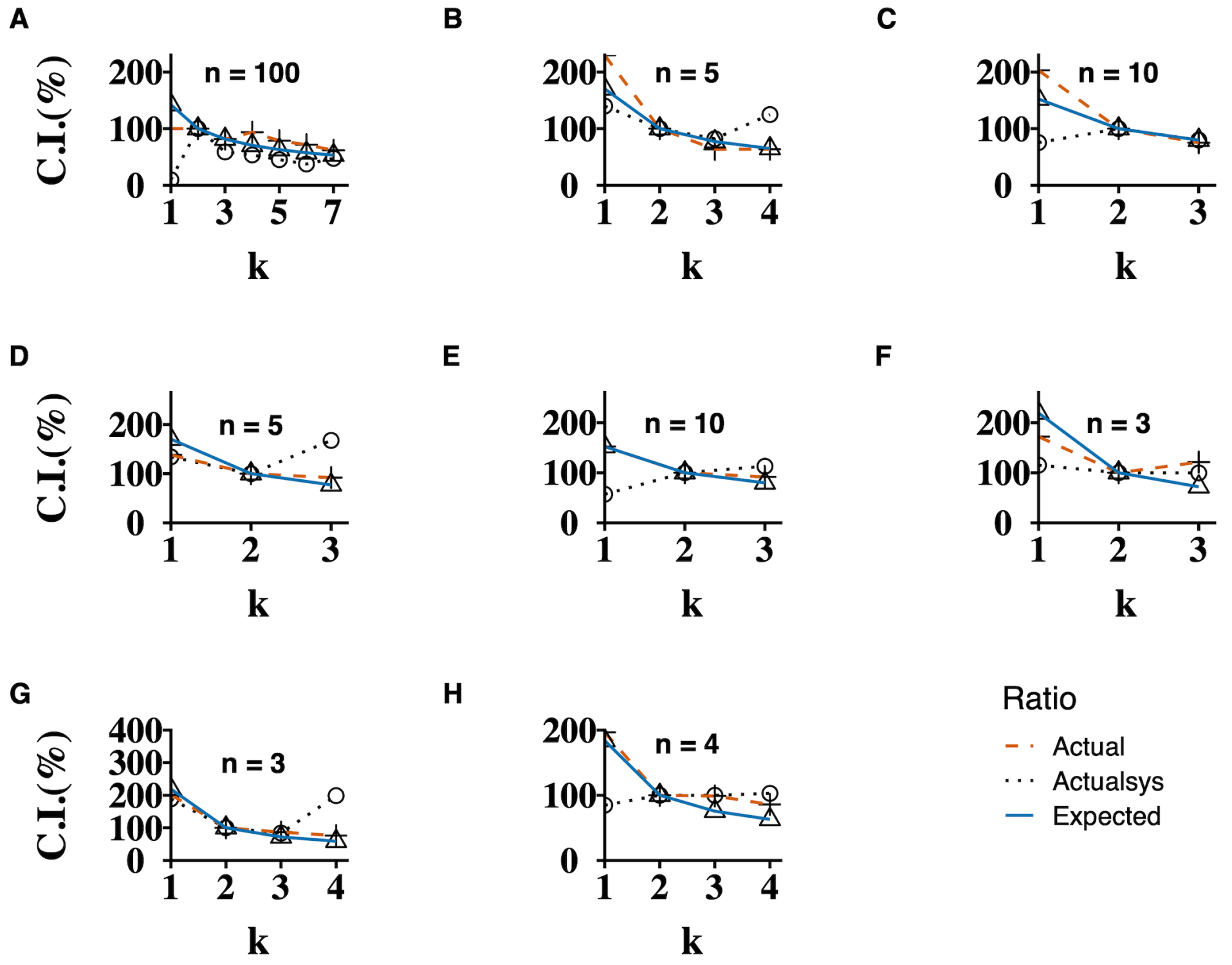

**Figure 4. Confidence interval (C.I.) ratios for real-world data.** C.I. widths (%; relative to that at k=2) versus number of replicates k, each with size n. Estimates from real-world data proxies without (Actual); or with (Actualsys) systematic error inclusion using variance among means (strictly: standard error of the grand mean estimated from the variance among the means); are shown. "Expected" gives the theoretical pattern with random error alone from a t distribution with replicates of size n. The proxy data has been treated by consecutively adding replicates to give pooled data in order to illustrate what sometimes happens in method replication. Data was extracted from articles: A. Tonk et al. [33]; B. Kichigina et al. [34]; C. Carlucci et al. [35]; D. Lawrence et al. [36]; E. Zhang et al. [37]; F. Driver et al. [38]; G. Zhou et al. [39] and H. unpublished data related to Clark et al. [40]. This graph has no Welch–Satterthwaite adjustment: Figs S7.B1 and S7.B2 show almost no effect on Actual curves, but there are substantial effects on Actualsys curves (Fig S7.C). Full details of extraction methods given in Supplemental files S7 and S8.

## Real-world-proxy data for batch-replication-CI intervals.

To demonstrate that theoretical-CI patterns sometimes apply to real-world data with global/method triplication, proxy data was extracted from protocol-report articles (Fig 4) or unpublished data (Supplemental_file_S8) and manipulated (Methods).

Firstly, with random error alone ("Actual", "NOsys") with CI real-data estimates, in all cases (except Fig 4F) lines roughly followed theoretical "expected" decreases (t-distribution-random-error alone). (Fig 4 has replicate random-error differences; in the idealized analysis below assumed constant.) Descriptive comparisons of actual versus expected CIs (file_S11): With A_Tonk the Actual (NOsys) pattern was significantly different from expected (k1:7; Fig S11A.1). However, the five informative points lay within the 25% additive tolerance band (S11:Section_8.1) and the k2:6 chunk (Fig S11A.2) was not statistically significantly different from expected.

All other Actual (NOsys) curves were not statistically-significantly different from expected and roughly followed expected curves: file_S11 graphs are reasonably convincing in five cases (Figs S11:B and D-E and G-H). Additionally, 14/15 informative points lay within 25% additive bands around expected; all apart from F_Driver_NOsys (file_S11:Section_8.1). (The argument this is simply because estimates are forced to scale with 1/k does not hold as can be seen from Fig 4F.)

With systematic-error estimates added (Actualsys; SYS; using variance among means: strictly: standard error of the grand mean estimated from the variance among the means), CI effects were often dominated by systematic-errors (>2× random error in 4/8 cases):

mean(systematic error)/mean(random error) ratios (file_S7:TABLE_S7.T16) were: A_Tonk ratio=45.4; B_Kichigina ratio=1.39; C_Carlucci ratio=2.04; D_Lawrence ratio=0.750; E_Zhang ratio=3.81; F_Driver ratio=0.531; G_Zhou ratio=1.31; H_Clark ratio=2.95.

With systematic error included: in 2/3 cases when n≥10 (Figs 4A and 4C, but not 4E) the lines including systematic error (Actualsys) decreased after k=2 (as predicted from idealized curves). GAM models were fitted from k=2 onwards (k=1 inclusion not appropriate) and needed a minimum of 3 points, so models could only be applied with total number of replicates ≥4. This gave four possible comparisons: A_Tonk, B_Kichigina; G_Zhou and H_Clark (Figs S11AA-BB and GG-HH), the latter three not being significantly different from expected curves (S11_Section_5.1; hardly surprising with only three points). With the (more important) tolerance-band descriptions: A_Tonk, B_Kichigina, C_Carlucci, G_Zhou and H_Clark had at least one actual (SYS) value within the additive (or multiplicative) tolerance bands (file_S11:Section_8.2); for A_Tonk and C_Carlucci: all informative points; overall 9/15 informative points in additive-tolerance bands.

Of importance, A_Tonk data was dominated by systematic error (45.4× random error) and with the tolerance-band descriptions, all five informative points for (SYS) A_Tonk fell within the 25% additive-tolerance band around expected (S11:Section_8.2). In other words, despite overwhelming systematic error, the SYS model behaved as expected as far as the additive-tolerance band was concerned but not according to GAM models which showed statistically-significant differences which are here regarded as less important: the graphs show slight differences in the slopes for actualsys versus expected (not due to Bessel correction use/non-use; this reflects small-sample sds declining more shallowly than root-mean-square expected curves and diminishes as k increases; Figs S11AA.1, AA.2).

The sensitivity analysis with Welch-Satterthwaite adjustment showed that the adjustment had almost no effect on the Actual (NOsys) curves but a substantial one on Actualsys, because when the systematic-error variance greatly exceeds the sampling variance the combined Welch degrees of freedom fall towards k−1 at k=2, inflating t_crit and hence the k=2 anchor.

## Minor studies.

Triplicated 1980s/1990s plant-science studies (n=100; Supplemental file_S2) were not analyzed in depth. As with physics examples (file_S6), inherent bias was likely substantial (results not discussed further).

## Lowest-level "subject" numbers.

Protocol N3, and lowest-level "subject" numbers, were not the present-article's main focus. However, two articles had n=2 (NoR); seven articles n=3 (N3); and two articles had n=4 (NoR).

From manipulated data in file_S11:Section 2: D_Lawrence (La); E_Zhang (Zh) with means treated as individual datapoints: At k=2 (values: La: 6.2, 5.3; Zh: 20.0, 23.5) interpoint SDs (ISDs) were $\mathrm{ISD}_{\mathrm{La},k=2} = 0.636$ and $\mathrm{ISD}_{\mathrm{Zh},k=2} = 2.47$, with relatively huge associated CIs (file_S11). At k=3 (La: 6.2, 5.3, 14.2; Zh: 20.0, 23.5, 27.2) ISDs were comparable at $\mathrm{ISD}_{\mathrm{La},k=3} = 4.90$ and $\mathrm{ISD}_{\mathrm{Zh},k=3} = 3.60$. However, normalized-Greenwood statistics (measuring datapoint-gap variances[42]) showed: $\mathrm{Gw}_{\mathrm{La},k=3} = 0.636$ (skewed; potentially warning about systematic errors) and $\mathrm{Gw}_{\mathrm{Zh},k=3} = 0.001$ (showing symmetry around the mean).

# Discussion.

## Protocol proportions.

Overall, 70% of studies had some form of replication. It is also clear that, in 2017, a large proportion (48%) of plant-science-research groups realized that triplication+ was important for part/entire studies (Fig 2). Despite huge pressures from limited resources, many had decided to triplicate+ at least part of their study before publication. In triplicated studies/part-studies, 1/3 budget/part-budget went into new hypothesis-generation testing and 2/3 into another activity, namely replication (providing no new knowledge concerning a new hypothesis). Presumably, many assumed that a non-replicated result was false until shown otherwise. Additionally, the number of studies with exact triplication was substantially and statistically-significantly increased, 2.65×, over expected; with a large decrease from triplication (25%) to quadruplication (5%).

Overall, replication-level reporting was weak/opaque. Compulsory replication-protocol Supplemental-flow charts could be recommended and, hopefully, definitions provided might be useful for future assessments of the state of a science through time. No 2017 replication-error-protocol studies (possibly indicating increased use of statistical tests), but one meta-analysis, were found.

The definitions of the protocols provided resolve numerous complications, and should, with minor modifications, be able to be used in the assessment of any science.

## Experimental/association-study type.

Experimental/association-study distinction is important for causal inference (and is also affected by the predominance of unknown systematic errors). However, causal decisions are separate from protocol decisions and no apparent association was detected (Fig 3): non-replicated-association study proportion not significantly higher than experimental. Theoretically, the replication of association studies could be assumed to be even more important than for experiments. For illustration see Fig 5, where combined-unknown-potential confounders are assumed to have approximately twice the opportunity to affect main parameters (P/Q) in association studies (whereas in experiments mainly outcome Q is affected; providing the causal-inference basis).

Theoretically, plant scientists might have considered publishing non-replicated associations via an alternative rationale as follows: (1) observational-bias risk might be great; (2) preference might have been to wait for future replicates from other researchers (with different sets of observational biases) and then meta-analyze for true/false hypothesis determination. Fig 3 provides no evidence this reasoning was followed.

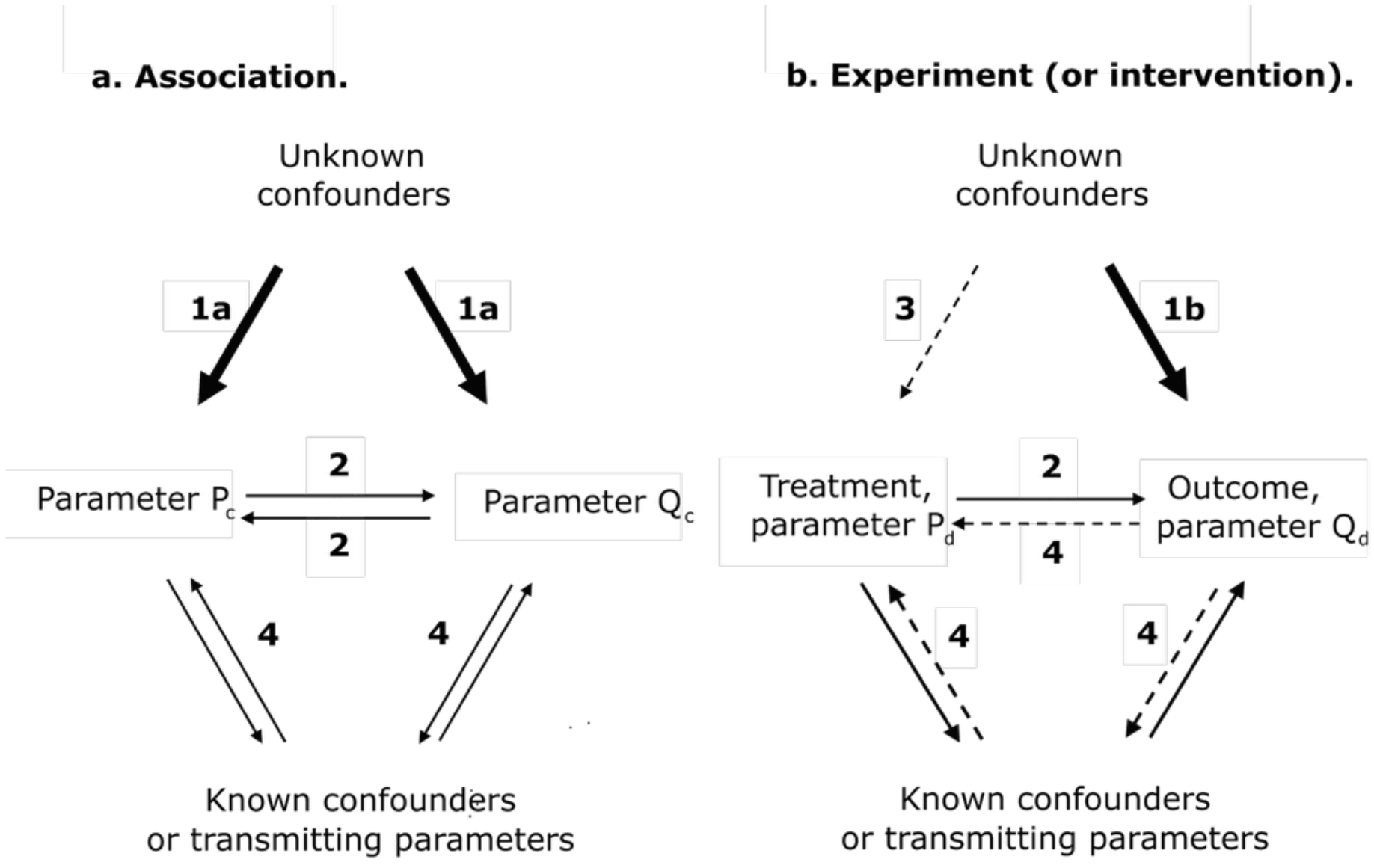

**Figure 5. Weighted directed graphs for a. association, and b. experimental, studies.** All arrows show possible, often unknown, causal effects, in the sense that, potentially, all can have effects in the direction indicated. Arrows 2 potentially contribute to true positive effects. Arrows 4 might also contribute to true-positive effects via transmitting parameters. All non-2 arrows might contribute potential systematic-error effects and, therefore, potential false positives. Relative thicknesses of lines are intended to represent the likelihood of effects: the weight of line 1b is estimated from Freedman et al. 2015[11] for medical science preclinical studies (weights of lines 1a assumed to be similar). Dashed lines 4 represent unlikely reverse effects which only occur under rare circumstances. Line 3 is where unknown confounders can affect the choice of sample treated. Known confounders are often countered by elimination or use as adjustors in statistical tests. (Parameters P and Q can correspond to those in Fig 1.)

## Meta-analyses.

One meta-analysis was found in the protocol report, not excluding extensive use: CI-upper limit for 1/(220 articles) is 6 (file_S7:Section_6) giving a reasonable average of 220/6=37 articles per meta-analysis. However, it seems from results that initial hypothesis falsification took place via other protocols; not disqualifying possible subsequent meta-analysis (as replicated studies can be included) [45–49].

## Lowest-level "subject" numbers.

While not the focus of the present article, (non-batch replication) lowest-level-"subject" numbers are used here to provide further explanation as to why (batch) triplication should be preferred. Non-replicated-study frequencies for (n=2)/(n=3)/(n=4) were 1/7/2, respectively (S7:Section_6). Real-world data were used with means interpreted as individual datapoints (file_S11): sometimes n=3 SDs were similar but normalized-Greenwood statistics (Gw) showed (descriptive) data-shape differences. The point being that Gw and other shape statistics (e.g.; generalized additive (GAM) models used later for dispersion), are invalid for n=2 (with only one gap). At n=3/triplication, shape can provide useful information, potentially concerning systematic errors and n=2 and n=3 provide qualitatively different information.

## Real-world-proxy data for batch-replication-CI intervals.

The proxy-data descriptive comparisons of actual (Actual; NOsys) versus expected CI patterns (GAM models; tolerance bands) showed that, in most cases, expected CIs were followed reasonably closely if only random error was considered.

With systematic-error estimates (Actualsys; SYS) included using standard error of the grand mean, systematic-error effects often dominated over random error. In one example (A_Tonk), the mean(systematic error)/mean(random error) ratio was 45.4×, but even so the actual CI pattern followed that expected closely according to the additive-tolerance band (14/15 informative points) and in five examples actualsys values fell within tolerance bands (9/15 informative points in additive-tolerance bands).

## Limitations:

(1) The small number of protocol-report articles (100); although a power study indicated sufficiency to show triplication excess.
(2) Not much can be said about excluded journals: only 10 articles were analyzed per journal, saving time (protocol-report execution took considerable time). Although all plant journals (>IF1) were assessed, plant-science coverage was moderate because of missing (unavailable) journals.
(3) Only one year was analyzed in the protocol report: 2017: this was a perfectly good twenty-first century representative year (see Introduction). Time trends could not be assessed (even though 1980s/1990s data are presented).
(4) To show broad applicability of (previously-known) square-root rules (see below), ideally many large, replicated, individual datasets (rarely published) should be assessed. However, here one example (Fig 4A; n=100) showed that, at least sometimes, rules apply even with overwhelming systematic error.

## Systematic-error effects.

Presumably, if unknown confounders which contribute to real-world variability are a ubiquitous aspect of reality, the best way of determining whether results are replicable is by replication. Increasing sample size is regarded as less effective than the setting up of separate replicates because "resetting" unknown confounders is important. Some confusion arises with non-exact replicates with known confounders. With a positive then a negative result: just because known confounders are present does not preclude predominance of unknown confounders; providing the likely reason for a false positive (true-negative prevalence always assumed to vastly exceed false-negative prevalence across all hypotheses).

# The theoretical underpinning of (batch) triplication.

### a) False-positive elimination (the most important aspect).

Triplication reduces the chances that a false-positive result will advance a true-hypothesis claim. Triplication effectiveness can be presumed from allocated resources but quantitative false-positive-reduction rates are unknown in any science. Plant-science failures (false-positive results with subsequent negative results) are not efficiently recorded. However, data can be borrowed from medical-science preclinical studies ([11]; Fig 5): failure rate 50%:90%. Assuming 70% median-failure rate, duplications could give $0.7^2 = 49\%$; triplications $0.7^3 = 34\%$; quadruplications $0.7^4 = 24\%$; assuming replicate independence. This remains our best replication-effectiveness guesstimate concerning false-positive elimination: triplication giving 31% less false positives than duplication $((0.49 - 0.34)/0.49 = 0.31)$; quadruplication a further 20% less than duplication $((0.34 - 0.24)/0.49 = 0.20)$. Assuming these (rather tenuous) calculations roughly correct, triplication would bring false-positive rates well below 50%, not achieved by duplication: one theoretical reason to prefer triplication.

## b) True positives.

Unknown confounders are also likely to affect true-positive results. The following discussion might be fairly obvious to a statistician, but it is probably worth collating triplication information as this is difficult to find this in peer-reviewed literature. Additionally, note that although statistically described, (batch) replication itself is not a statistical problem (in the sense that it cannot effectively be replaced by statistical techniques or simulation, which is a major point of this article).

The underlying main-effect-parameter random error (t, normal or other distribution) variance will scale with 1/k[50]. By assuming systematic errors (variance) also scale with 1/k (which occurs at least sometimes; see Sokal and Rohlfs'[9] analysis-of-variance partitioning), Fig 6 represents (idealized) CIwidth decrease with replicate number.

## c) True positives: k=1:k=2.

With identical replicate variances and normally-distributed data, relative true k=1:k=2 CIwidths theoretically decrease from 1 to $\sqrt{1/2}$ (41.4% relative to k=2; Fig 6). However, this would only be measured with negligible systematic-error effects, thought uncommon. With replicate-systematic error the k=1:k=2 change in measured CIwidth is likely to increase, not decrease, because k=1 true CIwidths (accounting for replicate-systematic errors) cannot be estimated. This was actually demonstrated by k=1:k=2 increases from Fig 4 real-world data (4A,C,E,H) (note: for small n, first-replicate small variance might also occur with random error).

### d) True positives: k=2:k=3.

Fig 6 shows a large decrease between duplication and triplication $\left(1 - \sqrt{2/3} = 18.4\%\right)$, with/without systematic errors. With only random-error estimated, a similar decrease is likely to be commonly measured, and was shown by 7/8 real-world-data proxies (Fig 4: all except F). With large sample sizes, decreases might also be measured even with included systematic-error estimates (as seen in Figs 4A-C and 4G).

### e) True positives: k=3:k>4.

A smaller drop occurs from triplication to quadruplication (1 to $\sqrt{3/4}$ or, relative to k=2, $\sqrt{(2/3)} - \sqrt{(2/4)} = 10.9\%$; Fig 6/Fig 4A), meaning quadruplication efficiency (CIwidth reduction; value accuracy) is ~1/2 that for triplication; with random error alone or 1/k-scaling systematic errors. With systematic errors considered, for one real-world example (A_Tonk; systematic error 45.4× random error) all five informative points fell within a tolerance band.

Some systematic errors will not scale with 1/k, but theoretical CI patterns (Figs 6 and S7:Fig 7.B) seem to be a good assumption for some cases (main parameters with normal, log-normal, t or other distributions) for random error alone or 1/k-scaling systematic errors.

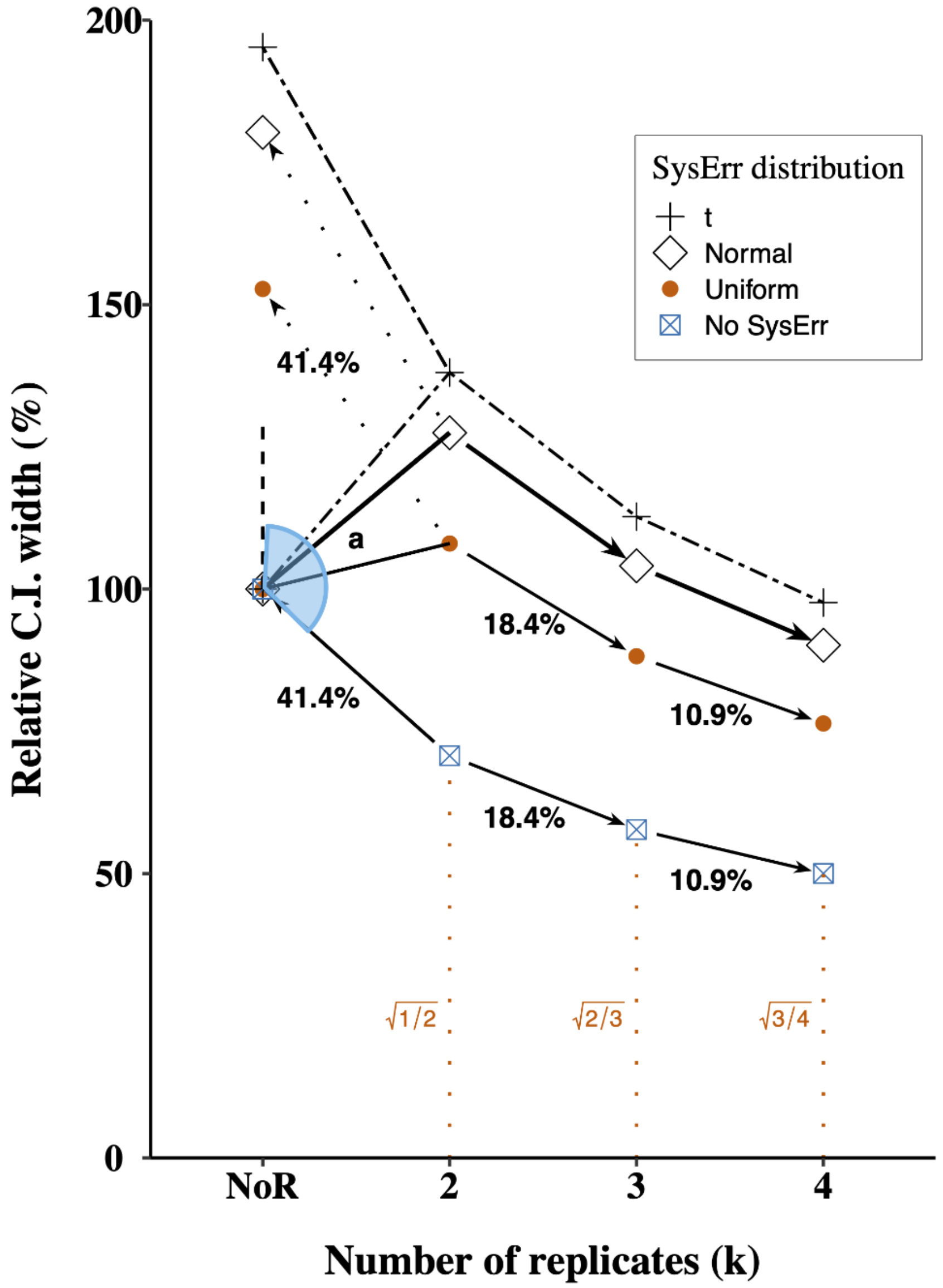

**Figure 6. Idealized theoretical confidence-interval (C.I.) ratios.** Theoretical patterns in confidence-interval widths with increase in replicate number (k; identical replicate random error), relative to no replication (NoR, k=1), with an underlying normal distribution either without systematic error(s) (SysErr) or with added systematic error with distribution: Normal, Uniform or t. At k=1 measured values (100%) are far below true widths due to replicate systematic errors which cannot be estimated from k=1 alone. Systematic-error sizes were chosen

aesthetically (Normal, t: 15%; Uniform: 12.5% of group means), but this only affects k=1 to k=2 directions ("a") which could vary according to the blue arc. Constants (%) show confidence-interval width changes relative to k=2, only depend on k (see text) and apply to all patterns displayed as arrows: k=2:k=1, 41.4% $= \sqrt{(2/1)} - 1$; k=2:k=3, 18.4% $= 1 - \sqrt{(2/3)}$; k=3:k=4, 10.9% $= \sqrt{(2/3)} - \sqrt{(2/4)}$. Square-root labels refer to: k=2 value relative to k=1, $\sqrt{(1/2)}$; k=3 value relative to k=2, $\sqrt{(2/3)}$; and k=4 relative to k=3, $\sqrt{(3/4)}$. Constants and square-root rules do not change with "a" direction or internal parameters for normal or uniform systematic error. A t-distribution systematic error (two-dash with no arrows; 10 degrees of freedom) is given for comparison, scaled exactly as for the normal-distribution systematic error: showing larger confidence-interval widths, and slightly steeper angles of decrease (see also file_S7:Fig S7.B). Parameters and coding given in file_S7.

## The basis for Triplication.

The basis for triplication can now be stated: (1) triplication gives a further reduction in false-positive likelihood over duplication; (2) non-replicated results give no inter-replicate systematic-error estimates; duplications do, but do not allow systematic-error-orderliness assessment (e.g., square-root rules (see below); mean shape): only a triple-result+ can give this information; (3) there is a large (CIwidth; accuracy) efficiency decrease between triplication and quadruplication.

## Square-root-rule mathematical relationships.

For normal distribution with random error (no systematic error):

$$\mathbf{CIwidth} = 2Z_{\mathrm{alpha/2}} \cdot \mathbf{SE_total}, \quad \text{where} \quad \mathbf{SE_total} = \frac{s}{\sqrt{kn}}, \qquad (4)$$

CIwidth = confidence interval width, $Z$ = standardized_normal quantile, alpha = significance level, SE_total = total standard error, $s$ = sample standard deviation, $k$ = number of replicates, $n$ = sample size.

For the ratio between two CI widths:

$$\frac{\mathbf{CIwidth_{k1}}}{\mathbf{CIwidth_{k2}}} = \frac{2Z_{\mathrm{alpha/2}} \cdot s/\sqrt{\mathrm{k1} \cdot n}}{2Z_{\mathrm{alpha/2}} \cdot s/\sqrt{\mathrm{k2} \cdot n}} \approx \frac{1/\sqrt{\mathrm{k1}}}{1/\sqrt{\mathrm{k2}}} = \sqrt{\mathrm{k2/k1}}, \qquad (5)$$

where: k1 = a replicate number, k2 = the next replicate number above k1. (Here: the equal-n, equal-s case is presented for clarity; ≈ indicates the ratio holds approximately when n and s differ.)[50]

What is perhaps surprising is that even with random error overwhelmed by systematic error whose variance contribution scales as 1/k (e.g., normally or uniformly distributed) results might be similar (this is not universally applicable and only applies to 1/k scaling; breadth of application should be determined empirically). For example, if

$$\mathbf{SE_total} \approx \sqrt{\left(\frac{\sum(\mathrm{mean} - \mathrm{global_mean})^2}{k(k-1)}\right)} \qquad (6)$$

where SE_total is here the standard error of the grand mean, then, assuming independent (zero-mean) per-replicate offsets (derivation in [9, 51] and file_S10):

$$\frac{\mathbf{CIwidth_{k1}}}{\mathbf{CIwidth_{k2}}} = \frac{2Z_{\mathrm{alpha/2}} \cdot \mathbf{SE_total_{k1}}}{2Z_{\mathrm{alpha/2}} \cdot \mathbf{SE_total_{k2}}} \approx \frac{1/\sqrt{\mathrm{k1}}}{1/\sqrt{\mathrm{k2}}} = \sqrt{\mathrm{k2/k1}} \qquad (7)$$

How closely real-world data fits square-root rules (equations 5,7) will partly depend on systematic/random-error ratios and/or if the t-distribution is appropriate. The latter results in even more pronounced k=2/k=3 and k=3/k=4 CIwidth ratios: the critical-t value depends on n,k, enhancing differences (file_S7:Fig S7.B).

With similar means, square-root rules simply reflect similar variance contributions from subsequent replicates. However, broad applicability of these rules remains, at present, a hypothesis: in 6/8 real-world proxy examples, CIwidths roughly followed expected curves with consideration of random error alone (file_S11:Section_8.1).

While relationships shown above might be well-known in some sciences, it is unclear whether the reasons for triplication as the optimal standard are widely known (not even in the plant sciences). In any case, while results here provide a small amount of evidence that sometimes real-world data follows these relationships, the extent of applicability still needs determination.

## Conclusions.

1. Triplication was introduced to counter known, unknown, or unforeseen, systematic errors contributing to real-world variability; and has been deployed since the nineteenth century.
2. Fisher's 1925 p<0.05 rule[19] was promptly incorporated into a p<0.05 three-times rule (triple-result protocol), countering both random and systematic error. In the twenty-first century, specifically in 2017, plant scientists used triple-result (11%) and triplication (25%) protocols extensively (including greater numbers of replicates: 17%,48%, respectively).
3. Triplication is the appropriate standard: (a) false positives reduce to acceptable levels (notably with triple results). With true positives: (b) there are qualitative differences between duplication/triplication information: duplication cannot show orderliness (shape); (c) there can be large efficiency decreases (in CIwidths, value accuracy; with other diminishing returns) between triplication/quadruplication, as shown by real-world-proxy data.
4. With random error alone, or 1/k-scaling systematic errors, square-root rules can, theoretically, show confidence-interval-width-ratio variation with k and this, for sufficient sample size n, sometimes applies to real data (as illustrated by one real-data example with dominant systematic error; six examples with only random error considered).
5. Triple-result protocols ($p<0.05$ three times), and triplication in general, could be recommended in any science for the variable construct discussed, as a minimum requirement to allow inductive inference for the purpose of further hypothesis generation. Indeed, the motto "no triplication, no full result" could be proposed.

# Acknowledgements.

One version of this article has been checked by an English Academic Editor. We would like to thank Prof. Mateusz Kurzawski (PUM, Szczecin, Poland), Prof. Nigel Halford (Rothamsted Research, UK) and John Clarkson (Bristol, UK) for discussions concerning replication; Prof. Raul Jimenez Tellado, University of Barcelona, Spain for discussions regarding systematic errors; and to Anna Sałacka, Agnieszka Boroń, Thierry van de Wetering, Krzysztof Safranow, Kazimierz Ciechanowski, Leszek Domański, and Andrzej Ciechanowicz for support. **Ethics approval and consent to participate:** Not applicable. **Consent for publication:** Not applicable. **Availability of data and materials:** All data are available in the main text or the supplementary materials. **Competing interests:** All authors declare they have no competing interests. **Funding:** there is no funding to report. **Authors' contributions:** JSCC contributed to data collection, concept/design, data analysis/interpretation, drafting the article; KP, KR and KS to data analysis/interpretation, critical revision of article.

# References.

1. Noreña J, Verde L, Jimenez R, Peña-Garay C, Gomez C. Cancelling out systematic uncertainties. Mon Not R Astron Soc. 2012;419:1040–50. https://doi.org/10.1111/j.1365-2966.2011.19761.x.

2. Byrnes JEK, Dee LE. Causal Inference With Observational Data and Unobserved Confounding Variables. Ecology Letters. 2025;28:e70023. https://doi.org/10.1111/ele.70023.

3. Philippe H, Poustka AJ, Chiodin M, Hoff KJ, Dessimoz C, Tomiczek B, et al. Mitigating Anticipated Effects of Systematic Errors Supports Sister-Group Relationship between Xenacoelomorpha and Ambulacraria. Current Biology. 2019;29:1818-1826.e6. https://doi.org/10.1016/j.cub.2019.04.009.

4. Zhao Y, Hao Z, Zhao C, Zhao J, Zhang J, Li Y, et al. A Novel Strategy for Large-Scale Metabolomics Study by Calibrating Gross and Systematic Errors in Gas Chromatography–Mass Spectrometry. Anal Chem. 2016;88:2234–42. https://doi.org/10.1021/acs.analchem.5b03912.

5. Natsidis P, Kapli P, Schiffer PH, Telford MJ. Systematic errors in orthology inference and their effects on evolutionary analyses. iScience. 2021;24:102110. https://doi.org/https://doi.org/10.1016/j.isci.2021.102110.

6. Nelson N, Stubbs CJ, Larson R, Cook DD. Measurement accuracy and uncertainty in plant biomechanics. Journal of Experimental Botany. 2019;70:3649–58. https://doi.org/10.1093/jxb/erz279.

7. Ingvarsson PK, Street NR. Association genetics of complex traits in plants. New Phytologist. 2011;189:909–22. https://doi.org/https://doi.org/10.1111/j.1469-8137.2010.03593.x.

8. Sokal RR, Rohlf FJ. Biometry : the principles and practice of statistics in biological research. San Francisco, CA, USA: W.H. Freeman; 1981.

9. Sokal R, Rohlf F. Biometry, 4th edn. New York, NY, USA: W.H. Freeman; 2012.

10. Valcin D, Jimenez R, Verde L, Bernal JL, Wandelt BD. The age of the Universe with globular clusters: reducing systematic uncertainties. Journal of Cosmology and Astroparticle Physics. 2021;2021:017. https://doi.org/10.1088/1475-7516/2021/08/017.

11. Freedman LP, Cockburn IM, Simcoe TS. The Economics of Reproducibility in Preclinical Research. PLOS Biology. 2015;13:e1002165. https://doi.org/10.1371/journal.pbio.1002165.

12. Clark JSC, Podsiadło K, Sobalska-Kwapis M, Marciniak B, Rydzewska K, Ciechanowicz A, et al. rs67047829 genotypes of ERV3-1/ZNF117 are associated with lower body mass index in the Polish population. Scientific Reports. 2023;13:17118. https://doi.org/10.1038/s41598-023-43323-3.

13. Escombe F. Germination of Seeds. I. The Vitality of Dormant and Germinating Seeds. Science Progress (1894-1898). 1897;6:585–608.

14. Peter A. Culturversuche mit “ruhenden” Samen.[In German]. Nachrichten von der Gesellschaft der Wissenschaften und der Georg-Augusts-Universität zu Göttingen. 1893;:673–91.

15. Bedford SA. Experimental Farm for Manitoba. In: Report of the Experimental Farms of the Dominion of Canada for the Year 1892. Ottawa: S. E. Dawson, Queen’s Printer; 1893. p. 189–225.

16. Keeble FW. The Hanging Foliage of Certain Tropical Trees. Annals of Botany. 1895;9:153–74.

17. Woods CD. On Sources of Error in Field Sampling and in the Chemical Analyses of Feeding Stuffs. In: Experiment Station Record, Volume 3: August 1891 to July 1892. Washington, D.C.: Government Printing Office; 1892. p. 379–86.

18. Schjerning H. The Culture of Barley for Malting Purposes. In: Experiment Station Record, Volume 6: July 1894 to June 1895. Washington, D.C.: Government Printing Office; 1895. p. 27–9.

19. Fisher RA. Statistical methods for research workers. Edinburgh, UK: Oliver and Boyd; 1925.

20. Fisher RA. The arrangement of field experiments. Journal of the Ministry of Agriculture. 1926;33:503–15.

21. Fisher, R. A. The statistical method in psychical research. Proceedings of the Society for Psychical Research. 1929;39:189–92.

22. Fisher RA. The design of experiments. 1935;:273 pp.

23. Preston RD, Bragg W. III—The Organization of the Cell Wall of the Conifer Tracheid. Philosophical Transactions of the Royal Society of London Series B, Biological Sciences. 1934;224:131–74. https://doi.org/10.1098/rstb.1934.0017.

24. Lamm R. Cytological Studies on Inbred Rye. Hereditas. 1936;22:217–40. https://doi.org/10.1111/j.1601-5223.1936.tb02649.x.

25. Thomas MD, Hill GR. The Continuous Measurement of Photosynthesis, Respiration, and Transpiration of Alfalfa and Wheat Growing under Field Conditions. Plant Physiology. 1937;12:285–307.

26. Benjamin DJ, Berger JO, Johannesson M, Nosek BA, Wagenmakers E-J, Berk R, et al. Redefine statistical significance. Nat Hum Behav. 2018;2:6–10. https://doi.org/10.1038/s41562-017-0189-z.

27. Randall, David, Welser, Christopher. The Irreproducibility Crisis of Modern Science: Causes, Consequences, and the Road to Reform. Princeton, NJ, USA: National Association of Scholars; 2018.

28. Lyons L. Open Statistical Issues in Particle Physics. The Annals of Applied Statistics. 2008;2:887–915.

29. Lyons L, Wardle N. Statistical issues in searches for new phenomena in High Energy Physics. J Phys G: Nucl Part Phys. 2018;45:033001. https://doi.org/10.1088/1361-6471/aa9408.

30. Fisher RA. Statistical methods for research workers. ed. 4. Edinburgh and London. 1932;:307 pp.

31. Brown MB. 400: A Method for Combining Non-Independent, One-Sided Tests of Significance. Biometrics. 1975;31:987–92.

32. Kost JT, McDermott MP. Combining dependent *P*-values. Statistics & Probability Letters. 2002;60:183–90. https://doi.org/10.1016/S0167-7152(02)00310-3.

33. Tonk FA, Kaya E, İştipliler D, İlker E, Turanlı F, Tosun M, et al. Identification of resistance to Eurygaster integriceps Put. on some bread wheat genotypes. Journal of Applied Botany and Food Quality. 2017;90:52–7. https://doi.org/10.5073/JABFQ.2017.090.008.

34. Kichigina NE, Puhalsky JV, Shaposhnikov AI, Azarova TS, Makarova NM, Loskutov SI, et al. Aluminum exclusion from root zone and maintenance of nutrient uptake are principal mechanisms of Al tolerance in Pisum sativum L. Physiol Mol Biol Plants. 2017;23:851–63. https://doi.org/10.1007/s12298-017-0469-0.

35. Carlucci A, Lops F, Mostert L, Halleen F, Raimondo M. Occurrence fungi causing black foot on young grapevines and nursery rootstock plants in Italy. Phytopathologia Mediterranea. 2017;56:10–39. https://doi.org/10.14601/Phytopathol_Mediterr-18769.

36. Lawrence BA, Lishawa SC, Hurst N, Castillo BT, Tuchman NC. Wetland invasion by *Typha* × *glauca* increases soil methane emissions. Aquatic Botany. 2017;137:80–7. https://doi.org/10.1016/j.aquabot.2016.11.012.

37. Zhang Y, Keller M. Discharge of surplus phloem water may be required for normal grape ripening. J Exp Bot. 2017;68:585–95. https://doi.org/10.1093/jxb/erw476.

38. Driver T, Trivedi DK, McIntosh OA, Dean AP, Goodacre R, Pittman JK. Two Glycerol-3-Phosphate Dehydrogenases from Chlamydomonas Have Distinct Roles in Lipid Metabolism. Plant Physiology. 2017;174:2083. https://doi.org/10.1104/pp.17.00491.

39. Zhou S, Jiang W, Long F, Cheng S, Yang W, Zhao Y, et al. Rice Homeodomain Protein WOX11 Recruits a Histone Acetyltransferase Complex to Establish Programs of Cell Proliferation of Crown Root Meristem. Plant Cell. 2017;29:1088–104. https://doi.org/10.1105/tpc.16.00908.

40. Clark J, Saunders D, Loeffler T, Hollomon DW. Purification of polysaccharide antigens from Pseudocercosporella herpotrichoides. German Phytomedical Society Series No 4. 1993;:313–23.

41. Ballico M. Limitations of the Welch-Satterthwaite approximation for measurement uncertainty calculations. Metrologia. 2000;37:61. https://doi.org/10.1088/0026-1394/37/1/8.

42. Greenwood M. The Statistical Study of Infectious Diseases. Journal of the Royal Statistical Society. 1946;109:85–110. https://doi.org/10.2307/2981176.

43. R Core Team. R: A Language and Environment for Statistical Computing. Vienna, Austria: R Foundation for Statistical Computing; 2026. https://doi.org/10.32614/R.manuals.

44. Cohen J. Statistical Power Analysis for the Behavioral Sciences. 2nd edition. New York: Routledge; 2013. https://doi.org/10.4324/9780203771587.

45. Scammacca N, Roberts G, Stuebing KK. Meta-Analysis With Complex Research Designs: Dealing With Dependence From Multiple Measures and Multiple Group Comparisons. Rev Educ Res. 2014;84:328–64. https://doi.org/10.3102/0034654313500826.

46. Gleser LJ, Olkin I, others. Stochastically dependent effect sizes. The handbook of research synthesis and meta-analysis. 2009;2:357–76.

47. Koricheva J, Gurevitch J. Uses and misuses of meta-analysis in plant ecology. Journal of Ecology. 2014;102:828–44. https://doi.org/https://doi.org/10.1111/1365-2745.12224.

48. Krupnik TJ, Andersson JA, Rusinamhodzi L, Corbeels M, Shennan C, Gerard B. Does size matter ? A critical review of meta-analysis in Agronomy. Exp Agric. 2019;55:200–29. https://doi.org/10.1017/S0014479719000012.

49. Rest JS, Wilkins O, Yuan W, Purugganan MD, Gurevitch J. Meta-analysis and meta-regression of transcriptomic responses to water stress in Arabidopsis. The Plant Journal. 2016;85:548–60. https://doi.org/https://doi.org/10.1111/tpj.13124.

50. Moore DS, McCabe GP, Craig BA. Introduction to the Practice of Statistics. New York, NY, USA: W.H. Freeman; 2009.

51. Montgomery DC. Design and analysis of experiments. 10th edition. Hoboken, NJ, USA: John Wiley; 2017.

# Supporting information captions.

**S1_Supplemental_File_S1_Definitions.docx:** Definitions of scientific terms used in the manuscript.

**S2_Supplemental_File_S2_Plant_Science_Examples_1980-1999.xlsx:** Examples (n=100) of plant science studies from 1980 to 1999 containing triplication (those with triplication: Global (replication-error) protocols fully processed).

**S3_Supplemental_File_S3_PRISMA.docx:** Prisma flow diagram.

**S4_Supplemental_File_S4_Plant_journal_GRID_2017.xlsx:** Grid of all 2017 plant science journals initially assessed for the Protocol Report.

**S5_Supplemental_File_S5_PROTOCOL_REPORT_PLANTS_2017.xlsx:** The Protocol Report contains all plant science articles included in protocol assessment, plus those articles excluded from the same journals.

**S6_Supplemental_File_S6_Physics_Examples.xlsx:** Examples of physics studies (n=25) giving explicit systematic error estimates.

**S7_Supplemental_File_S7_CODING.docx:** Main coding file for manuscript statistics and graphs.

**S8_Supplemental_File_S8_Plant_Science_CI.xlsx:** List of original data sources used for real-world confidence interval changes with increasing number of replicates.

**S9_Supplemental_File_S9_Collated.xlsx:** S5 altered to allow direct extraction of Protocol Report data.

**S10_derivation_eq6_to_eq7.docx:** Mathematical derivation of equation 7.

**S11_Supplemental_File_S11_R_GAM_CODING_H.docx:** Coding for statistics and graphs resulting from analyses of real-world data from S8 (via S7).

**S12_Supplemental_File_S12_articles from JOURNALS_EXCLUDED.xlsx:** Articles from journals which were excluded entirely (with reasons).

**S13_Supplemental_File_S13_heuristic_for_Protocol_Report.docx:** Heuristic to allow future protocol report analyses.